\documentclass[a4paper,12pt]{article}
\usepackage{graphicx}	
\usepackage{amsmath}	
\usepackage{amssymb}	
\usepackage{subfigure}
\usepackage{sidecap}
\usepackage{tikz}
\usepackage{multirow}
\usepackage{authblk}
\usepackage{hyperref}
\usepackage{txfonts}
\usepackage{newtxtext,newtxmath}
\date{}
\providecommand{\keywords}[1]{\textbf{\textit{Keywords:}} #1}

\begin{document} 
\title{\textbf{On the Interpretation of the Aharonov-Bohm Effect}}

\author{Jay Solanki}
\affil{Sardar Vallabhbhai National Institute of Technology,\\ Surat - 395007, Gujarat, India\\ E-mail: jay565109@gmail.com }

\maketitle

\begin{abstract}
The Aharonov-Bohm (A-B) effect showed that the phase of electron wave pattern could be changed by the excluded electromagnetic field, the region where electromagnetic field is zero. This apparent non-local effect has been explained by mainly two salient interpretations called "the interpretation of electromagnetic potentials" and "the interpretation of interaction energy". In this paper the author reviews and investigates both the interpretations and analysis their differences. Then the author clarifies in  details and extend the approach of interaction energy interpretation and argue that as quantum mechanics involved energy of quantum systems and electromagnetic energy contained in electromagnetic field, the natural and more physically acceptable interpretation of the effect should be "the interpretation of interaction energy".


\end{abstract}

\section{Introduction}
In 1959, Aharonov and Bohm investigated that the phase of an electron wave could be changed by excluded electromagnetic field, the region where electromagnetic field is zero.\cite{PhysRev.115.485, PhysRev.123.1511} Thus, it appears that the phase of electron wave is affected by remote electromagnetic field through action at a distance. This seems to be a challenging problem because field theory was discovered to avoid action at distance. Most of the research in this topic is aimed to solve this non-locality problem. Thus, in order to overcome from this non-local interpretation of the effect, Aharonov and Bohm suggested that this effect could be produced by electromagnetic potentials that are non-zero in the region where electrons are passes through. This effect is known as A-B effect. There are two A-B effects namely the electric A-B effect and the magnetic A-B effect. In this paper the author only focus on the magnetic A-B effect. 
\\ In order to prove the magnetic A-B effect Aharonov and Bohm proposed an experiment. In the proposed experiment they suggested to use solenoid with axis in the vertical direction. Theoretically infinitely long solenoid can constrain magnetic field inside it. Practically very long solenoid can be used such that magnetic field outside the solenoid can be neglected or instead of solenoid, toroidal magnets can be used. They suggested to use a coherent electron beam that splits into two parts, each one going through each side of the solenoid. Both the splitted electron beam should be brought together behind the solenoid to create interference pattern due to quantum mechanical effects between electron beams. According to quantum mechanics electron beams behave like waves, creating interference pattern when bought together. Moreover in Aharonov-Bohm paper they predicted that there should be the phase change in interference pattern produced by electron beams due to the presence of the solenoid. This magnetic A-B effect has been extensively studied theoretically and practically by various authors.
\cite{2015FrPhy..10..358W, doi:10.1119/1.1976266,  Fe_nberg_1963, inbook, PhysRevA.86.040101, PhysRevLett.56.792, PhysRevLett.5.3, Jaklevic1965MacroscopicQI, epub3511, article,doi:10.1142/S0217732306020962, doi:10.1142/S0217732399001541, doi:10.1142/S0217732301004297, doi:10.1142/S0217732311035766, doi:10.1142/S0217732394001805}
There are numerous experiment carried out to prove the magnetic A-B effect.\cite{PhysRevLett.56.792, PhysRevLett.5.3, Jaklevic1965MacroscopicQI, epub3511, article} The remarkable experiment carried out by Tonomura et al. is widely consider as the firm evidence of the A-B effect.\cite{PhysRevLett.56.792} 
\\  In A-B effect it appears that magnetic field remotely affects electron wave packets asserting the action at distance. However the field theory was originally constructed to avoid action at distance. This problem has received substantial interest. To tackled this problem there are mainly two approaches have been suggested to avoid action at distance. These two approaches or interpretations are known as "the interpretation of electromagnetic potentials"\cite{PhysRev.115.485, PhysRev.123.1511, inbook} and "the interpretation of interaction energy"\cite{2015FrPhy..10..358W, Fe_nberg_1963, article2, article1}. However it is unclear that which interpretation is the correct one. The possible reason behind the confusion could be that most of the studies presume one interpretation as basics of the detailed study. In particular, no study, to the author knowledge has considered direct implementation of the fundamental laws of electromagnetism to solve the problem. Thus in this paper the author first review both the interpretation in sections (2) and (3). In section (4) the author undergone a rethinking of the problem and developed a method by direct implementation of the laws of electromagnetism to solve the problem and to find physically acceptable interpretation. The most important advantage of this method is that it is implemented by direct use of fundamental laws of electromagnetism, so that it provides more natural and simplified explanation of the A-B effect. In section (5) the author briefly review experimental work carried out to verify A-B effect and to understand correct reason behind the A-B effect, followed by discussion in section (6).

\section{The Electromagnetic Potential Interpretation}
By considering suggested Aharonov-Bohm experiment discussed in introduction, in this section we briefly review explanation of the A-B effect given in terms of the electromagnetic potential interpretation by many authors.\cite{PhysRev.115.485, PhysRev.123.1511, inbook}  In the theory of electromagnetism, potentials $ \phi $ and $ \mathbf{A} $  are used to describe electromagnetic fields $ \mathbf{E} $ and $ \mathbf{B} $. In classical electromagnetism $ \mathbf{E} $ and $ \mathbf{\textbf{B}} $ are considered to be more fundamental then potentials $ \phi $ and $ \mathbf{A} $. However because of  potential $ \phi $ and $ \mathbf{A} $ appears in Schrodinger equation (1), some physicist asserted that  potential $ \phi $ and $ \mathbf{A} $ are more fundamental than fields $ \mathbf{E} $ and $ \mathbf{B} $. 
\begin{equation}
    i\hbar\frac{\partial\Psi}{\partial t} = \left[ \frac{(\hat{P} - q\mathbf{A})^2}{2m} + q\phi \right]\Psi
\end{equation}
Where $ \Psi $ is the wave function and $ \hat{P} = -i\hbar\nabla$  is momentum operator.
Now as of Hamiltonian, Lagrangian of the particle in electromagnetic field is also depends on electromagnetic potentials. In order to find phase shift it is useful to carry out analysis in terms of propagator rather than Schrodinger equation. The propagator is given by\cite{https://doi.org/10.1002/zamm.19590390514,lennard-jones_1931}
\begin{equation}
    K(\mathbf{r}_f,t_f:\mathbf{r}_i,t_i) = \int[D\mathbf{r}]e^{iS[\mathbf{r}(t)]/\hbar}
\end{equation}

Where the action of for electron can be written as
\begin{equation}
    S[\mathbf{r}(t)] = \int dt\left( \frac{1}{2}m\mathbf{v}^2 + \frac{e}{c}\mathbf{A}\cdot\frac{d\mathbf{r}}{dt}  \right)
\end{equation}
 In the magnetic Aharonov-Bohm experiment solenoid produces only static magnetic field confined inside it. Thus there is only vector potential exist and scalar potential is zero everywhere due to solenoid. Thus action only depends on vector potential produced by solenoid in equation (3). In the above action second term is responsible for phase shift in electron wave packets. The phase difference $\frac{(S1-S2)}{\hbar}$ between both electron beams becomes,\cite{PhysRev.115.485}
\begin{equation}
    \frac{(S1-S2)}{\hbar} = \frac{e}{\hbar c}\int \mathbf{A}\cdot d \mathbf{r} = \frac{e}{\hbar c} \Phi_0 
\end{equation}
Where $ \Phi_0 $ is magnetic flux enclosed by splitted electron beams.
\\Equation (4) describes phase difference between electron beams due to presence of the current carrying solenoid. If there would be not current carrying solenoid then vector potential will be zero and phase difference between the electron beam is zero. Now because of the direct appearance of the electromagnetic potential in above calculation, many physicists considers electromagnetic potentials as more fundamental than electromagnetic field in quantum mechanics. Thus the electromagnetic potential interpretation asserts that the observed phenomenon of the change of the phase of electron beam is caused by the electromagnetic potential because the electromagnetic field is zero in that region.

\section{The Interaction Energy Interpretation}
In this section the author briefly review explanation of A-B effect by the interaction energy interpretation. This interpretation asserts that the phase change of electron beam is produced by the energy of the interaction between electron beam's electromagnetic field and solenoid's electromagnetic field. Let's assume that long solenoid produces a static magnetic field $ B_1(\mathbf{r}) $ inside it and electron beam produces magnetic field $ B_2(\mathbf{r}) $ around it. The total magnetic field in the space is given \begin{equation}
    B(\mathbf{r}) = B_1(\mathbf{r}) + B_2(\mathbf{r})
\end{equation}                         

The energy of the total magnetic field is given by,
\begin{equation}
    U = \frac{1}{2\mu_0}\int \mathbf{B}_1 ^2(\mathbf{r})d^3r + \frac{1}{2\mu_0}\int \mathbf{B}_2 ^2(\mathbf{r})d^3r + \frac{1}{\mu_0}\int \mathbf{B}_1(\mathbf{r})\cdot\mathbf{B}_2(\mathbf{r})d^3r
\end{equation}

The first two terms are the energy of magnetic field produced by solenoid and electron beam respectively. The third term is the interaction energy between the magnetic fields $\mathbf{B}_1(\mathbf{r})$ and $\mathbf{B}_2(\mathbf{r})$.\cite{2015FrPhy..10..358W} It was shown that Interaction energy is related to the vector potential $\mathbf{A}_1$ describing the magnetic field $\mathbf{B}_2(\mathbf{r})$ produced by the solenoid.\cite{Rui_Feng_2009}
\begin{equation}
    \int \frac{1}{\mu_0}\mathbf{B}_1(\mathbf{r})\cdot\mathbf{B}_2(\mathbf{r})d^3r = \mathbf{A}_1(\mathbf{r})\cdot q\mathbf{v}
\end{equation}
where $\mathbf{A}_1(\mathbf{r)}$ is given by
\begin{equation}
    \mathbf{A}_1(\mathbf{r}) = \frac{1}{4\pi}\int \frac{\mathbf{B}_1(\mathbf{r'}) \times (\mathbf{r} - \mathbf{r'})}{|\mathbf{r} - \mathbf{r'}|^3}d^3r'
\end{equation}

From equation (7) we can conclude that although interaction energy and formula for phase change of electron can be described in terms of vector potential, the magnetic A-B effect is actually caused by the interaction energy of magnetic fields $\mathbf{B}_1$ and $\mathbf{B}_2$. 

\section{Physically Acceptable Interpretation according to Electromagnetism}
In this section the author consider the fundamental laws of electromagnetism to determine the physically acceptable and correct interpretation of A-B effect. One widely acceptable law of electromagnetism asserts that the energy of the electrodynamic system is stored in electromagnetic field.\cite{jackson_classical_1999, 1964flp..book.....F} Now Quantum mechanics is formulated in terms of energy of quantum systems, rather than in terms of force. Thus, when we describe quantum mechanical effects of electrodynamic system, we have to consider electromagnetic energy of that system. In magnetic A-B effect we have to consider magnetic energy of the system of solenoid and electron beam to describe quantum mechanical effect of phase change of the electron beam. Now mathematically magnetic energy of the system can be written in terms of current density and magnetic potential as follow,\cite{jackson_classical_1999, 1964flp..book.....F}
\begin{equation}
U = \frac{1}{2} \int \mathbf{j}\cdot\mathbf{A}\ dV
\end{equation}
Here appeared magnetic potential is just a mathematical function to calculate magnetic energy and it's not a real physical field. The main reason to not consider electromagnetic potentials as real physical quantity is their non uniqueness. There are also some other disadvantages are there in considering electromagnetic potentials as real physical quantities as discussed in reference \cite{2015FrPhy..10..358W}. We can also write magnetic energy in terms of physical magnetic field as follow,

\begin{equation}
U = \frac{1}{2\mu_0} \int \mathbf{B} \cdot \mathbf{B} \ dV
\end{equation}

Here in equation (10) magnetic energy appeared in terms of magnetic field. That asserts that equation (9) is just another mathematical way to calculate magnetic energy. Now electromagnetic energy is considered to be stored in electromagnetic field. This becomes more acceptable when we write energy conservation law for electromagnetic system.

\begin{equation}
  -\frac{\partial u}{\partial t} =  \nabla \cdot \mathbf{S} + \mathbf{E} \cdot \mathbf{j} 
\end{equation}
Where energy density u and energy flux $\mathbf{S}$ are given by, 
\begin{equation}
u = \frac{\epsilon_0}{2}  \mathbf{E} \cdot \mathbf{E} + \frac{1}{2 \mu_0}  \mathbf{B} \cdot \mathbf{B} 
\end{equation}

\begin{equation}
\mathbf{S} = \frac{1}{\mu_0}  \mathbf{E} \times \mathbf{B}
\end{equation}

Here equation (11) assets that the loss of electromagnetic energy per unit volume is equal to sum of the energy flux flowing out through the surface boundary of that unit volume and work done by the fields on the source within the volume. here energy density and energy flux is naturally comes out in terms of electromagnetic field, which supports idea that electromagnetic energy is stored in electromagnetic field. Thus, we should consider electromagnetic potentials as just a mathematical tool to calculate energy of electromagnetic field. 
\\ Now in Aharonov-Bohm experiment, the quantum mechanical effect is produced by energy consideration of solenoid and electron beam because quantum mechanics deal with energy of quantum mechanical systems. In this electromagnetic system, energy is stored in electromagnetic field of electron beam and solenoid. Thus from electromagnetic law considered in this section and the interaction energy interpretation reviewed in the section (3), it becomes clear that quantum mechanical A-B effect is caused by the interaction energy between magnetic fields $\mathbf{B}_1$ and $\mathbf{B}_2$ as considered in the interaction energy interpretation. If we consider A-B effect in terms of electromagnetic potential as in electromagnetic potential interpretation, we should consider role of electromagnetic potential as just a mathematical tool rather than physical quantity. From the analysis presented in this section we can conclude that the interaction energy interpretation is more physically acceptable interpretation of A-B effect.

\section{Experimental Verification of A-B Effect }
There are many experiments are carried out and have been proposed to experimentally verify the A-B effect and to find the correct interpretation of the effect. The very famous experiment was carried out by Tonomura in 1986 to verify the effect and to understand the correct interpretation of the effect.\cite{PhysRevLett.56.792} But in most of the experiment the interaction of the electromagnetic fields and electromagnetic potential both exists simultaneously, making it impossible to find the correct reason behind the A-B effect. However in 1986 Tonomura conducted experiment with toroidal magnets covered with superconductors. In this experiment electron beam was splitted into two, each one going through inside the hole and out side of the toroidal magnet. They bought together and interference pattern was measured  by an interferogram. Now because of the observed interference pattern in the absence of electromagnetic energy interaction, it was believed that the cause of A-B effect are electromagnetic potentials. However by detailed analysis, it was found that superconductor cannot confine magnetic field produced by the electron beams. Thus electromagnetic energy interaction is still exist. Thus, also in reference \cite{2015FrPhy..10..358W} it is argued that if superconductor can confine magnetic field of the electron beam then interference will be same for all samples, proving that the correct reason behind the A-B effect is the electromagnetic field not potential. 

\section{Discussion}
In this paper, the author reviewed the Aharonov-Bohm effect and its two interpretations, namely "the interpretation of electromagnetic potentials" and "the interpretation of interaction energy". After reviewing both interpretations, the author implemented a more straightforward analysis of A-B effect to demystify the correct interpretation of the effect. In that analysis, the author considered the law of electromagnetism, which asserts that the energy of the electrodynamic system is stored in its electromagnetic field. As quantum mechanical effects are considered in terms of energy of quantum mechanical systems, the A-B effect should be considered due to the interaction energy of electromagnetic fields of the quantum mechanical system, in this case, the system of solenoid and electron beam. In electromagnetism, energy can be written both in terms of the electromagnetic field and electromagnetic potentials. However, energy written in terms of electromagnetic potentials is just a mathematical alternative that does not represent physical reality. The energy written in terms of the electromagnetic field represents physical reality as it leads to the fundamental energy conservation law for the electromagnetic system. Thus, electromagnetic potential in the equations of quantum mechanics should be thought of as just mathematical tools rather than physical quantity. From that analysis presented in section (4), the author supports the interaction energy interpretation of the A-B effect by providing a more straightforward analysis of the A-B effect considering the fundamental laws of electromagnetism. 

\section{Acknowledgement}
This work has been done independently. The author thanks BoseX, SVNIT research group members for fruitful discussions.

\nocite{Leus2013ThePE}
\nocite{ershkovich2013electromagnetic}
\nocite{Giuliani_2010}
\nocite{sym12122110}
\nocite{PhysRevLett.54.2469}
\nocite{doi:10.1080/00018738600101921}
\nocite{epub3511}
\nocite{Jaklevic1965MacroscopicQI}
\nocite{doi:10.1142/S021773230401549X}
\nocite{doi:10.1142/S0217732306020962}
\nocite{doi:10.1142/S0217732399001541}
\nocite{doi:10.1142/S0217732312500277}
\nocite{doi:10.1142/S0217732301004297}
\nocite{doi:10.1142/S0217732311035766}
\nocite{doi:10.1142/S0217732394001805}

\bibliographystyle{unsrt}
\bibliography{references.bib}

\end{document}